\documentclass[prd,aps,showpacs,preprintnumbers,amssymb]{revtex4}

\usepackage{color}
\usepackage{epsf}

\begin{document}
\title{%
\hfill{\normalsize\vbox{%
\hbox{}
 }}\\
{About electrodynamics, standard model and the quantization of the electrical charge}}

\author{Renata Jora
$^{\it \bf a}$~\footnote[2]{Email:
 rjora@theory.nipne.ro}}

\affiliation{$^{\bf \it a}$ National Institute of Physics and Nuclear Engineering PO Box MG-6, Bucharest-Magurele, Romania}

\date{\today}

\begin{abstract}
The quantization of the electrical charge in the electrodynamics and of the hypercharge in the standard model are imposed in the theory based not on theoretical arguments but on the experimental observations.  In this paper we propose a quantum consistency condition in a theory where Ward identities are respected that requires  the quantization of the charge within the framework of the theory without external impositions. This refers to the  renormalization conditions in the background gauge field method such that to  ensure a correct mathematical correspondence between the bare partition function and the renormalized one. Applied to the standard model of elementary particles our criterion together with the anomaly cancellation condition leads to the correct hypercharge assignment of all standard model fermions.

\end{abstract}
\maketitle

\section{Introduction}
 The standard model of elementary particles \cite{Glashow}-\cite{Hagen} is one of the most successful theories in physics with  a tremendous amount of experimental confirmation culminating with the discovery of the Higgs boson by the Atlas \cite{Atlas} and CMS \cite{CMS} experiments. However no indication of beyond of the standard model particles or interactions has emerged in almost a decade of LHC collecting data.  Issues like baryon asymmetry in the Universe, the nature and properties of neutrinos, dark matter or dark energy have remained largely unresolved.

 Some particular properties of the standard model had to be established phenomenologically in the absence of any confirmed UV completion of this theory. Among the facts confirmed experimentally with great accuracy but theoretically not justified is the quantization of the elementary charge.  In practice we know at least three instances when the quantization of the electrical charge is emergent in a theory.  The first is very old and stems from Dirac who showed that in a theory with a magnetic monopole the electrical charge is quantized.  The second is somehow more recent and  refers to the embedding of the abelian gauge group in a larger non-abelian one like $SU(5)$  such that the spontaneous breaking of the symmetry leads naturally to the abelian group.  Moreover the quantization of charge is a requirement in many QED lattice gauge theories \cite{K} which in general provide the only known non-perturbative regularization of a gauge theory.

In this paper we will explore a possibility never discussed in the literature to our knowledge, namely the fact that the quantization of the abelian charge can be an intrinsic property of the model (or derived within the model)  if some consistency condition between the bare and renormalized partition functions is imposed.

 In section II we will show that for simple electrodynamics such consistency condition requires that all charged fermions have the same absolute charge. This consistency condition is further discussed in section III.  In section IV we will apply the same procedure to the standard model fermions to obtain that this together with the cancellation of all anomalies associated with the hypercharge not only leads to the correct hypercharge assignment of all standard model fermions but also to a confirmation of these. Section V is dedicated to a general overview and discussion. The adequate conclusions are drawn.

\section{Quantization of the abelian charge}

We start from an abelian gauge theory, for simplicity denote this QED, with a set of fermions of different charges:
\begin{eqnarray}
{\cal L}=-\frac{1}{4}F^{\mu\nu}F_{\mu\nu}+\sum_f\bar{\Psi}_f(i\gamma^{\mu}\partial_{\mu}+gQ_f\gamma^{\mu}A_{\mu})\Psi_f.
\label{res6635529}
\end{eqnarray}

Next we consider the following change of variables:
\begin{eqnarray}
A_{\nu}\rightarrow A_{\nu}+\alpha B_{\nu},
\label{res647739}
\end{eqnarray}
where $\alpha$ is a constant parameter and $B_{\nu}$ is a background gauge field. The transformation in Eq. (\ref{res647739}) is first regarded just as a shift of variables and does not change the variable of integration in the partition function.
In terms of the new variable the Lagrangian will become:
\begin{eqnarray}
&&{\cal L}=-\frac{1}{4}F^{\mu\nu}F_{\mu\nu}+\sum_f\bar{\Psi}(i\gamma^{\mu}\partial_{\mu}+gQ_f\gamma^{\mu}A_{\mu})\Psi_f-
\nonumber\\
&&\frac{1}{2}\alpha F^{\mu\nu}B_{\mu\nu}-\frac{1}{4}\alpha^2B^{\mu\nu}B_{\mu\nu}+\alpha\sum_f\bar{\Psi}_fQ_fg\gamma^{\mu}\Psi_fB_{\mu}.
\label{trsnflagrr77466}
\end{eqnarray}
Now we consider two one particle incoming and outgoing fermion states which may be of any flavor in the model (for simplicity we denote them generically) with the same four momentum $p^{\mu}$ denoted by $|p'\rangle$ and $\langle p|$. We are interested in the matrix element:
\begin{eqnarray}
\langle p |\frac{\partial S}{\partial {\alpha}}|p'\rangle,
\label{matrelem76857}
\end{eqnarray}
where $S$ is the $S$ matrix. It is evident that:
\begin{eqnarray}
&&\langle p |\frac{\partial S}{\partial {\alpha}}|p'\rangle=\langle p |i\int d^4x [-\frac{1}{2} F^{\mu\nu}B_{\mu\nu}+\sum_f\bar{\Psi}_fQ_fg\gamma^{\mu}\Psi_fB_{\mu}]|p'\rangle_{\alpha=0}=
\nonumber\\
&&\langle p |p'\rangle \langle 0|\frac{\partial S}{\partial {\alpha}}|0 \rangle|_{\alpha=0}=0,
\label{resu775664}
\end{eqnarray}
because the whole thing  amounts to the equation of motion which is zero by the Dyson Schwinger approach. The following term in the matrix element in Eq. (\ref{resu775664}):
\begin{eqnarray}
\langle p|\int d^4 x \sum_f\bar{\Psi}_fQ_fg\gamma^{\mu}B_{\mu}\Psi_f|p'\rangle=\bar{u}(p)Qg\gamma^{\mu}B_{\mu}(p-p')u(p)=2p^{\mu}B_{\mu}(0)Qg
\label{res8857749}
\end{eqnarray}
can be easily computed because is just the definition of the electromagnetic vertex (for details see \cite{Weinberg}). Here the charge $Q$ corresponds to the external fermions and the contribution by other species of fermions is zero. This can be easily seen from the first term on the right hand side of the first line of Eq. (\ref{resu775664}) by simply applying the LSZ theorem.

The next step is to determine the contribution:
\begin{eqnarray}
\langle p |i\int d^4x [-\frac{1}{2} F^{\mu\nu}B_{\mu\nu}]|p'\rangle_{\alpha=0},
\label{contr7756647}
\end{eqnarray}
besides the obvious approach involving the transfer of the gauge boson states outside through the LSZ theorem.

For that let us start with the same matrix element for an arbitrary $\alpha$ namely:
\begin{eqnarray}
\langle p |i\int d^4x [-\frac{1}{2} F^{\mu\nu}B_{\mu\nu}]|p'\rangle.
\label{newmtsry65774}
\end{eqnarray}
We make a change of variables in the associated $S$ matrix $A_{\mu}=\frac{1}{g}A_{\mu}'$. Then there is a contribution coming from the jacobian of the change of variables that we denote by $J(g)$. We consider the matrix element:
\begin{eqnarray}
\langle p|S'|p'\rangle,
\label{res63554}
\end{eqnarray}
where $S'$ is the S matrix in the new variables, and differentiate with respect to $g$:
\begin{eqnarray}
&&\frac{\partial}{\partial g}\langle p|S'|p'\rangle=
\nonumber\\
&&\langle p|\frac{\partial S'}{\partial g}|p'\rangle=
\nonumber\\
&&\langle p|i\int d^4 x\Bigg[\frac{1}{2g^3}F^{\prime\mu\nu}F^{\prime}_{\mu\nu}+\frac{1}{2g^2}\alpha F^{\prime\mu\nu}B_{\mu\nu}\Bigg]|p'\rangle+\langle p|\frac{\partial J(g)}{\partial g}|p'\rangle.
\label{firstround65774}
\end{eqnarray}
Here we used the fact that the external states are independent of the coupling constant g. From Eq. (\ref{firstround65774}) one may derive:
\begin{eqnarray}
&&\langle p|i\int d^4x \alpha F^{\prime\mu\nu}B_{\mu\nu}|p'\rangle=
\nonumber\\
&&2g^2\Bigg[\frac{\partial}{\partial g}\langle p|S'|p'\rangle-[\langle p|i\int d^4x \frac{1}{2g^3}F^{\prime\mu\nu}F^{\prime}_{\mu\nu}|p'\rangle+\langle p|\frac{\partial J(g)}{\partial g}|p'\rangle]\Bigg],
\label{res663554}
\end{eqnarray}
or,
\begin{eqnarray}
&&\langle p| i\int d^4x F^{\prime\mu\nu}B_{\mu\nu}|p'\rangle=
\nonumber\\
&&\frac{2g^2}{\alpha}\Bigg[\frac{\partial}{\partial g}\langle p|S'|p'\rangle-[\langle p|i \int d^4x \frac{1}{2g^3}F^{\prime\mu\nu}F^{\prime}_{\mu\nu}|p'\rangle+\langle p|\frac{\partial J(g)}{\partial g}|p'\rangle]\Bigg].
\label{res663554}
\end{eqnarray}

Let us discuss each of the three terms on the right hand side of Eq. (\ref{res663554}). First we will give the results and then establish in what context these results are valid. The first term is:
\begin{eqnarray}
&&X_1(\alpha)=\frac{2g^2}{\alpha}\frac{\partial}{\partial g}\langle p|S'|p'\rangle=
\nonumber\\
&&\frac{2g^2}{\alpha}\langle p|p'\rangle\frac{\partial}{\partial g}\langle 0|S'(\alpha)|0\rangle.
\label{firsterm657774}
\end{eqnarray}
For the massless fermions with the same momenta the normalization of states is given by $\langle p|p'\rangle=2E_{\vec{p}}(2\pi)^3\delta(\vec{p}-\vec{p'})$  (where for simplicity we omitted the spin) and thus is independent of the species of fermions.  This term is then flavor independent.

The third term on the right hand side of Eq. (\ref{res663554}) yields:
\begin{eqnarray}
X_3(\alpha)=\frac{2g^2}{\alpha}\langle p|\frac{J(g)}{\partial g}| p'\rangle=
\frac{2g^2}{\alpha}\langle p |p'\rangle \langle 0|\frac{\partial J(g)}{\partial g}|0\rangle,
\label{thridtermresweew}
\end{eqnarray}
and consequently this term is also independent of the fermion flavor.

The second term on the right hand side of Eq. (\ref{res663554}) is:
\begin{eqnarray}
&&-\frac{2g^2}{\alpha}\langle p|i \int d^4x \frac{1}{2g^3}F^{\prime\mu\nu}F^{\prime}_{\mu\nu}|p'\rangle=
\nonumber\\
&&=-\frac{2g^2}{\alpha}\langle p|i\int d^4x [\frac{1}{2g^3}G^{\mu\nu}G_{\mu\nu}-\frac{\alpha}{g^2}G^{\mu\nu}B_{\mu\nu}]|p'\rangle=
\nonumber\\
&&-\frac{2g^2}{\alpha}\langle p|i\int d^4x \frac{1}{2g^3}G^{\mu\nu}G_{\mu\nu}|p'\rangle+2\langle p|i\int d^4x G^{\mu\nu}B_{\mu\nu}|p'\rangle.
\label{secondterm4677756}
\end{eqnarray}
Here in the second and third lines we performed the change of variable in the $S$ matrix $G^{\nu}=A^{\prime\nu}+\alpha gB^{\nu}$.  The matrix element will thus be taken on states with $\alpha=0$ and thus one can safely redefine $G_{\mu\nu}=F^{\prime}_{\mu\nu}$ (Note that there is an extra contribution of order $\alpha^2$ which is zero in the limit considered). We add  Eqs. (\ref{firsterm657774}), (\ref{thridtermresweew}) and (\ref{secondterm4677756}) and use Eq. (\ref{res663554}) to obtain:
\begin{eqnarray}
&&-\langle p| i \int d^4 xF^{\prime\mu\nu}B_{\mu\nu}|p'\rangle|_{\alpha=0}=
\nonumber\\
&&\lim_{\alpha\rightarrow 0}\Bigg[X_1(\alpha)+X_3(\alpha)-\frac{2g^2}{\alpha}\langle p|i\int d^4 x\frac{1}{2g^3}F^{\prime\mu\nu}F^{\prime}_{\mu\nu}|p'\rangle|_{\alpha=0}\Bigg]=
\nonumber\\
&&\lim_{\alpha\rightarrow 0}\Bigg[X_1(\alpha)+X_3(\alpha)-\frac{2g^2}{\alpha}\langle p|p'\rangle\langle 0|\frac{\partial S}{\partial g}|0\rangle|_{\alpha=0}\Bigg]=X(B_{\mu},p).
\label{finalterm6577474}
\end{eqnarray}
Thus the matrix element in Eq. (\ref{res663554}) is independent of the fermion species. Here in the final result of the right hand side we made a change of variable back to $A_{\mu}$ in the partition function.

From Eq. (\ref{resu775664}), (\ref{res8857749}) and (\ref{finalterm6577474}) we determine:
\begin{eqnarray}
|2p^{\mu}B_{\mu}(0)Qg|=\frac{1}{2}|X(B_{\mu},p)|,
\label{resfinal664553}
\end{eqnarray}
where the right hand side is independent of the fermion species and the left hand side contains a particular charge $Q$. We used the absolute value because the sign of  the limit in Eq. (\ref{finalterm6577474}) is undetermined depending on the particular sign of the parameter $\alpha$ around zero.  The conclusion that imposes itself is that in order for the theory to be consistent the renormalized charge of all fermion species in the model is not only quantized but can take only the values $\pm1$. Note that in the discussion we omitted beforehand the particles that do not interact with the gauge field as they are decoupled from the theory. Our result is not valid if the theory contains other interactions of the gauge fields. It is thus absolutely necessary for the gauge field to interact only with the charged fermions.

It seems that such a drastic result cannot be true. Indeed there is an underlying assumption substantiating it. It refers to all equations starting from Eq. (\ref{res663554}) to Eq.(\ref{finalterm6577474}). If $\alpha$ is zero in the $S$ matrix beforehand then the renormalized theory follows the standard rules. However if $\alpha\neq0$ as it is in the matrix elements in the equations mentioned then we have a theory with a background gauge field. Then the renormalization of the theory is different, the renormalization constants are also different. All  the operations we performed are in general not consistent in the renormalized perturbation theory. Note that in our calculations we assumed that one can trade the bare theory with the renormalized one without altering the final results. Here we shall make the following crucial assumption that regulates all the possible inconsistencies that may appear in our approach: that the parameter $\alpha$ gets renormalized such as to cancel the renormalization of the coupling constant and that the theory in the presence or absence of the background gauge field is in its entirety renormalized exactly in the same way such that the two methods may be related through a shift of the gauge variable. This amounts to requiring that:  $\alpha g =\alpha_rg_r$ where the index $r$ denotes the renormalized quantity  and all other renormalization constants (including that for the gauge parameter) are the same for the theory with or without the background gauge field. We claim that this assumption is necessary for the bare theory to be at least mathematical compatible to the renormalized one. Then the background gauge theory is renormalized exactly as the theory with $B_{\mu}=0$ and all procedures and operations we performed are consistent both in the bare and renormalized theory. In conclusion our condition of renormalization is equivalent to the quantization of charge of an abelian theory, fact verified for all phenomenological theories.

 Before ending this section it is necessary to highlight how the presence of fermion masses may change our result. There are  a few  places where an eventual fermion mass may appear: in the right hand side of Eq. (\ref{res8857749}) and in Eqs.(\ref{firsterm657774}), (\ref{thridtermresweew}) and (\ref{finalterm6577474}). In the last three equations  the masses appear in the matrix element $\langle p|p'\rangle=2E_{\vec{p}}(2\pi)^3\delta(\vec{p}-\vec{p'})$ which factorizes in $X(B_{\mu},p)$. Then in order to prove our point we just need to consider the momentum $\vec{p}$  zero. Then $E_p=m$ in the term $X(B_{\mu},p)$ and the left hand side of Eq. (\ref{resfinal664553}) will also contain $m$. Consequently  the factor $m$ cancels on the left hand side and right hand side of Eq. (\ref{resfinal664553}) and the corresponding charge remains universal and independent of the particular mass.

 \section{Discussion of the consistency requirement}

Let us see in detail what our consistency condition at the end of section II means for a gauge fixed abelian gauge theory with one fermion flavor. The Lagrangian of interest is:
\begin{eqnarray}
{\cal L}=-\frac{1}{4}F^{\mu\nu}F_{\mu\nu}+\bar{\Psi}i\gamma^{\mu}D_{\mu}\Psi-\frac{1}{2\xi}(\partial_{\mu}A^{\mu})^2.
\label{QDagain65774}
\end{eqnarray}

The background gauge field method amounts to a change in the action $A_{\mu}\rightarrow A_{\mu}+B_{\mu}$ but not in the gauge fixing term:
\begin{eqnarray}
{\cal L}(B_{\mu}) =-\frac{1}{4}[\partial^{\mu}(A_{\nu}+B_{\nu})-\partial^{\nu}(A_{\mu}+B_{\mu})]^2 -\frac{1}{2\xi}(\partial_{\mu}A^{\mu})^2+\bar{\Psi}i\gamma^{\mu}(\partial_{\mu}-ie(A_{\mu}+B_{\mu})\Psi.
\label{newQEdlagtrbacj5664}
\end{eqnarray}
It is necessary for the gauge fixing term to have the form in Eq. (\ref{newQEdlagtrbacj5664}) and not one that includes the background gauge field. To see that note that the Lagrangian in Eq. (\ref{newQEdlagtrbacj5664}) is invariant under the following transformation of the fields:
\begin{eqnarray}
&&B_{\mu}(x)\rightarrow B_{\mu}(x)-\frac{1}{e}\partial_{\mu}\alpha(x)
\nonumber\\
&&A_{\mu}(x)\rightarrow A_{\mu}(x)
\nonumber\\
&&\Psi(x)\rightarrow \exp[i\alpha(x)]\Psi(x)
\nonumber\\
&&\bar{\Psi}(x)\rightarrow \exp[-i\alpha(x)]\bar{\Psi}(x).
\label{tarnsofmr6646655}
\end{eqnarray}
The invariance under the transformations in Eq. (\ref{tarnsofmr6646655}) ensures that the background  gauge field effective action has the correct gauge properties.

It is known that in QED  $Z_1=Z_2$ in each order of perturbation theory where $Z_1$ is the renormalization of the fermion kinetic term and $Z_2$ is the renormalization constant of the fermion interaction term. Moreover $Z_eZ_A^{1/2}=1$ where $Z_e$ is the renormalization constant of the charge and $Z_A$ is the renormalization constant of the gauge field kinetic term.  Then in the absence or the presence of the background gauge field the renormalization conditions for both the background gauge field and the quantum fluctuations may be  the same:
\begin{eqnarray}
(A^{\mu}+B^{\mu})_0=Z_A^{1/2}(A^{\mu}_r+B^{\mu}_r).
\label{renormcond77566464}
\end{eqnarray}
Here the subscript $0$ denotes the bare fields whereas the subscript $r$ refers to the renormalized ones.

In an arbitrary field  theory with a generic  field $A$  and with a  source dependent partition function $Z[J]$ the connected Green functions are generated by taking derivative with respect to $J$ of the generating functional $W[J]=-i\ln Z[J]$.  In the presence of the background gauge field \cite{Abbott} the generating functional becomes:
\begin{eqnarray}
W[J,B]=-i\ln Z[J,B],
\label{genfunc657}
\end{eqnarray}
where $Z[J,B]$ is calculated  with  the substitution $A\rightarrow A+B$.  One may write;
\begin{eqnarray}
W[J,B]=W[J]-JA,
\label{genfunctryy56}
\end{eqnarray}
which is obtained by making back the shift $A\rightarrow A-B$.

All Green functions of interest and the full perturbation theory then are equivalent in the presence or the absence of the background gauge field because one can always make  the shift back to the original field  variable. In the case of a gauge field theory this statement is not valid anymore. If we make the change back to the original variable ($A_{\mu}\rightarrow A_{\mu}-B_{\mu}$) in Eq. (\ref{newQEdlagtrbacj5664}) one obtains:
\begin{eqnarray}
{\cal L}=-\frac{1}{4}F^{\mu\nu}F_{\mu\nu}+\bar{\Psi}i\gamma^{\mu}D_{\mu}\Psi-\frac{1}{2\xi}(\partial_{\mu}(A^{\mu}-B_{\mu}))^2,
\label{lagrchang664775}
\end{eqnarray}
which is different from the original Lagrangian in Eq. (\ref{QDagain65774}) due to the gauge fixing term. This means that the theory in the background gauge field has different properties  although the renormalization conditions are the same (except for the gauge parameter). But the total equivalence between the two theories was the necessary constraint stated at the end of section II that provided the  quantization of the charge. By simple inspection of the two Lagrangians one notes  that our consistency condition  leads to the derived condition $\xi\rightarrow \infty$.  This corresponds to the unitary gauge.  The unitary gauge does not contain unphysical degrees of freedom but it is problematic even for a massive gauge field because lacks the off-shell renormalizability. The latter point means that the off-shell Green function contain ultraviolet divergences that cannot be removed  by regular renormalization procedures \cite{Sonoda1}, \cite{Sonoda2} . Thus our consistency condition  lead to an old and not fully solved problem of the gauge quantum theory.  There are studies ( as for example \cite{Sonoda2}) that suggests that the unitary gauge can be used safely for higher order loop calculations. These are based on the fact that in the coordinate space the off-shell Green function may possess the property of multiplicative renormalizability thus making the unitary gauge more amenable. However the unitary gauge remains an open issue that needs to be addressed in its full complexity and is beyond the scope of our work here. Since in principle any gauge choice even problematic may lead to physical results that are not evident in other gauge choices our arguments for the quantization of charge in section II remain valid.

\section{Quantization of the hypercharge}
In this section we will apply the results of section II to the standard model fermions.  Since after spontaneous symmetry breaking the fermions  gain masses and also the electromagnetic field interacts with other gauge bosons we cannot apply our procedure to the electromagnetic charge.  However our method can be easily implemented for the hypercharge group $U(1)_Y$ before spontaneous symmetry breaking.

We start by writing the couplings of the standard model fermions with the hypercharge field through the gauge covariant derivatives. For one generation of fermions this reads:
\begin{eqnarray}
D_{\mu}\Psi_L=(\partial_{\mu}-ig\vec{T}\vec{A}_{\mu}-ig'\frac{Y_L}{2}Y_{\mu})\Psi_L,
\label{left477663}
\end{eqnarray}
for the left handed states, and,
\begin{eqnarray}
D_{\mu}\Psi_R=(\partial_{\mu}+ig'\frac{Y_R}{2}Y_{\mu})\Psi_R,
\label{right4664}
\end{eqnarray}
for the right handed ones.
Here we denoted the hypercharge gauge boson by $Y_{\mu}$ and $Y_L$ and $Y_R$ are the hypercharge assignments of the left handed and right handed states respectively. Of course  $Y_L$ and $Y_R$ takes different values for each species of fermions.

We will apply equations (\ref{resu775664}), (\ref{res8857749}) and (\ref{resfinal664553}) directly noting that the left hand side of Eq. (\ref{resfinal664553}) has the same expression and the right hand side will be modified according to:
\begin{eqnarray}
&&\langle p|\int d^4 x [\bar{\Psi}_Lg'\frac{Y_L}{2}\gamma^{\mu}B_{\mu}\Psi_L-\bar{\Psi}_Rg'\frac{Y_R}{2}B_{\mu}\Psi_R]|p'\rangle=
\nonumber\\
&&\langle p|\int d^4 x [\bar{\Psi}\gamma^{\mu}\Psi \frac{g'}{2}(Y_L-Y_R)B_{\mu}-\bar{\Psi}\gamma^{\mu}\gamma^5\Psi\frac{g'}{2}(Y_L+Y_R)B_{\mu}]|p'\rangle=
\nonumber\\
&&2p^{\mu}B_{\mu}(0)(Y_L-Y_R).
\label{res102938}
\end{eqnarray}
Here we used the fact that the second term on the second line of Eq. (\ref{res102938}) brings no contribution as the associated vertex is zero (this can also be verified by direct calculations).

We make the following notations: $Y_{Le}$ and $Y_{Re}$ for the hypercharges associated to the left and right handed charged leptons respectively, $Y_{Lu}$ and $Y_{Ru}$ for the hypercharges associated to the left and right handed up quarks and $Y_{Ld}$ and $Y_{Rd}$ for the hypercharges associated to the left handed and right handed down quarks.

Then from Eq. (\ref{resfinal664553})with the result in Eq. (\ref{res102938}) implemented on the left hand side we derive:
\begin{eqnarray}
|Y_{Le}-Y_{Re}|=|Y_{Lu}-Y_{Ru}|=|Y_{Ld}-Y_{Rd}|.
\label{res746635}
\end{eqnarray}
We will make the choice:
\begin{eqnarray}
Y_{Le}-Y_{Re}=-(Y_{Lu}-Y_{Ru})=Y_{Ld}-Y_{Rd},
\label{res}
\end{eqnarray}
for further use (without loss of generality).

Note that the correct definition of the hypercharges in the standard model without making any phenomenological assumption is:
\begin{eqnarray}
&&Y_{Le}=x(Q_e-T_{e})
\nonumber\\
&&Y_{Re}=xQ_e
\nonumber\\
&&Y_{Lu}=y(Q_u-T_{u})=Y_{Ld}=z(Q_d-T_d)
\nonumber\\
&&Y_{Ru}=yQ_u
\nonumber\\
&&Y_{Rd}=z(Q_d-T_d),
\label{res773663con}
\end{eqnarray}
where $T$ with a subscript is the third component of isospin.
By itself the definition in Eq. (\ref{res773663con}) does not insure the quantization of the hypercharge or  the electric charge and one needs beyond of the standard model physics to explain charge quantization.

However the constraint in Eq (\ref{res746635}) leads to $x=y=z$ which is all one needs for the quantization of the charges besides the cancellation of the anomalies.

Before discussing the anomalies however let us obtain another relation in our method.

We apply again Eq. (\ref{resfinal664553}) this time with the matrix $\gamma^5$ sandwiched between the external states. We discuss first the right hand side. The $\gamma^5$ matrix brings down anomalies in the $S$ matrix when it is associated with a delta function. We briefly review the procedure (for details see \cite{Srednicki}). We start with:
\begin{eqnarray}
\delta(0)=\lim_{x\rightarrow y}\delta(x-y)=\int \frac{d^4k}{(2\pi)^4}\exp[\frac{[i\gamma^{\mu}D_{\mu}]^2}{M^2}]\exp[i(x-y)].
\label{anom758844}
\end{eqnarray}
Here $D_{\mu}$ is the covariant derivative associated with the fermion corresponding to the external state. We ignore all other gauge interaction and for the quarks we assign definite color index without summation over the color numbers at any stage. Since for the right hand side the $S$ matrix is defined in the background gauge field method we have:
\begin{eqnarray}
D_{\mu}=\partial_{\mu}-i\frac{g'}{2}Y_L\frac{1-\gamma^5}{2}B_{\mu}+i\frac{g'}{2}Y_R\frac{1+\gamma^5}{2}B_{\mu},
\label{cocb66477}
\end{eqnarray}
where $B_{\mu}$ is the background gauge field.

Eq. (\ref{anom758844}) may be rewritten as:
\begin{eqnarray}
\delta(0)=M^4\int \frac{d^4k}{(2\pi)^4}\exp[ikM(x-y)]\exp[-k^2]\exp\Bigg[\frac{2ikD}{M}+\frac{D^2}{M^2}+\frac{g'}{2}(Y_L\frac{1+\gamma^5}{2}+Y_R\frac{1-\gamma^5}{2})S^{\mu\nu}F_{\mu\nu}{M^2}\Bigg].
\label{revbnn56}
\end{eqnarray}
Here we made a change of variables $\frac{k}{M}\rightarrow k$ and $S_{\mu\nu}=\frac{i}{2}[\gamma^{\mu},\gamma^{\nu}]$. Since ${\rm Tr}[S^{\mu\nu}]=0$ and ${\rm Tr} [S^{\mu\nu}\gamma^5]=0$ the $\gamma^5$ factor in front will bring down only the square of the last operator in the exponent.  Consequently,
\begin{eqnarray}
&&{\rm Tr}[\delta(0)\gamma^5]=
\nonumber\\
&&\frac{1}{8}g^{\prime 2}\int \frac{d^4k}{(2\pi)^4}\exp[-k^2]B_{\mu\nu}B^{\rho\sigma}{\rm Tr}\Bigg[\gamma^5S^{\mu\nu}S_{\rho\sigma}[Y_L^2\frac{1+\gamma^5}{2}+Y_R^2\frac{1-\gamma^5}{2}]\Bigg].
\label{finalres64553}
\end{eqnarray}
It is not necessary to compute the coefficients of the right hand side of Eq. (\ref{finalres64553}) in detail. The result can be written simply as:
\begin{eqnarray}
{\rm Tr} [\delta(0)\gamma^5]=a(Y_L^2+Y_R^2)Y^{\mu\nu}\tilde{Y}_{\mu\nu}+b[Y_L^2-Y_R^2]Y^{\mu\nu}Y_{\mu\nu},
\label{finalresuuuty7}
\end{eqnarray}
where $a$ and $b$ are coefficients universal for all fermion states and $\tilde{Y}_{\mu\nu}$ is the dual hypercharge  tensor.

The left hand side of Eq. (\ref{resfinal664553}) with a $\gamma^5$ matrix sandwiched between the external states gives the following vertex:
\begin{eqnarray}
&&\langle p|\gamma^5\frac{g'}{2}[\bar{\Psi}\gamma^{\mu}\Psi(Y_L-Y_R)B_{\mu}+\bar{\Psi}\gamma^{\mu}\gamma^5\Psi(Y_L+Y_R)B_{\mu}|p'\rangle=
\nonumber\\
&&-2p^{\mu}B_{\mu}(0)\frac{g'}{2}(Y_L+Y_R).
\label{axial647776}
\end{eqnarray}
This result is obvious if one observes that the presence of the $\gamma^5$ matrix transforms the vector term into an axial one and the axial one into a vector one.

Collecting the results form Eqs. (\ref{finalresuuuty7}) and (\ref{axial647776}) one obtains:
\begin{eqnarray}
A(Y_L^2+Y_R^2)+B(Y_L^2-Y_R^2)=C(Y_L+Y_R),
\label{eq7788}
\end{eqnarray}
where the coefficients $A$, $B$, $C$ depend on the background gauge field but not on the fermion species and Eq. (\ref{eq7788}) is valid for each type of fermions, namely,
\begin{eqnarray}
&&A\frac{Y_{Le}^2+Y_{Re}^2}{Y_{Le}+Y_{Re}}+B1=C
\nonumber\\
&&A\frac{Y_{Lu}^2+Y_{Ru}^2}{Y_{Lu}+Y_{Ru}}-B1=C
\nonumber\\
&&A\frac{Y_{Ld}^2+Y_{Rd}^2}{Y_{Ld}+Y_{Rd}}+B1=C,
\label{res774553662}
\end{eqnarray}
where $B_1=B(Y_{Le}-Y_{Re})$.
Eq. (\ref{res774553662}) leads immediately to:
\begin{eqnarray}
\frac{Y_{Le}^2+Y_{Re}^2}{Y_{Le}+Y_{Re}}=\frac{Y_{Ld}^2+Y_{Rd}^2}{Y_{Ld}+Y_{Rd}}.
\label{fin4601999}
\end{eqnarray}

It is important to note at this point that what we derived here makes sense only if the Ward identities are respected at each order in perturbation theory. This amounts to asking for the cancellation of all anomalies associated with the hypercharge gauge boson. In what follows we will briefly discuss those (since they are well known we give the final result of the calculation):

a) The anomaly of a $U(1)_Y$ gauge boson with two $SU(3)$ gauge bosons require:
\begin{eqnarray}
Y_{Lu}-Y_{Ru}+Y_{Ld}-Y_{Rd}=0.
\label{first6664}
\end{eqnarray}
This is respected already by the constraint we obtained in Eq. (\ref{res746635}) together with the choice in Eq. (\ref{res}).

b) The anomaly of a $U(1)_Y$ gauge boson and two $SU(2)_L$ gauge bosons requires:
\begin{eqnarray}
Y_{Le}+3Y_{Lu}=0.
\label{sec36677}
\end{eqnarray}

c) The gravitational anomaly with a $U(1)_Y$ gauge boson requires:
\begin{eqnarray}
2Y_{Le}-Y_{Re}=0.
\label{third6774664}
\end{eqnarray}

d) Finally the anomaly of three $U(1)_Y$ gauge bosons requires:
\begin{eqnarray}
2Y_{Le}^3-Y_{Re}^3+6Y_{Lu}^3-3Y_{Ru}^3-3Y_{Rd}^3=0.
\label{fourth567744}
\end{eqnarray}

The system of equations (\ref{sec36677}), (\ref{third6774664}) and (\ref{fourth567744}) together with the constraint obtained in this approach in Eq. (\ref{res}) can be solved and leads to:
\begin{eqnarray}
&&Y_{Lu}=-\frac{1}{3}Y_{Le}
\nonumber\\
&&Y_{Re}=2Y_{Le},
\label{resfinal7748}
\end{eqnarray}
where some of the equations are redundant.
All other hypercharges can be derived from these again from Eq. (\ref{res773663con}) in terms of $Y_{Le}$ showing that indeed the hypercharge is quantized. In the absence of the constraint in Eq. (\ref{res}) one would have four anomalies with three valid equations (Eq. (\ref{third6774664}) can be used only for the determination of the electric charge of the charged leptons) and five unknowns $x$, $y$, $z$ and $Y_{Le}$, $Y_{Lu}$. The final results will depend on two parameters and would not ensure the quantization of the charge.

Finally the last result obtained here in Eq. (\ref{fin4601999}) is a check of our method and it is verified for the solution in Eq. (\ref{resfinal7748}).

\section{Conclusions}

In general in electrodynamics or even in the standard model the experimentally observed quantization of charge is implemented by hand. Although this is a feature observed in most theories of this type there is not apparently any intrinsic theoretical reason within  the theory to justify the quantization of  the abelian charge. Here we propose a consistency condition  of the quantum theory that upon satisfying leads to the quantization of charge.  This condition is directly connected with the background gauge field method which is often used in physics. There is a known fact that the properties of a system  are different in the presence of a  nontrivial vacuum or  a source.
But the background gauge field is just the source associated with the gauge field. In general the system is treated differently in the presence of a background gauge field (as opposed to the absence of it) and also the renormalization conditions and constant are different.

Here we imposed the following consistency condition that in principle relates the bare and renormalized partition functions: the background gauge field gets renormalized exactly as the quantum fluctuations and all renormalizations constants and conditions are the same for both cases.  The only difference between the quantum fluctuation and the background gauge field is that the latter is not integrated in the partition function. We thus claim that promoting the background gauge field from a classical field to a quantum one that gets renormalized leads naturally and correctly to the quantization of charge.  In general this condition can be implemented easily in non-gauge theories. In gauge theories there are difficulties that arise due to the gauge fixing condition. Our consistency condition at the end of section II then is equivalent with the requirement that not only all fields and coupling constants are renormalized in the same way in the presence or absence of the background gauge field but also that the unitary gauge is implemented at all steps.  We discussed this in detail in section III. The unitary gauge has advantages at one loop but disadvantages at higher orders because of the lack of off-shell renormalizability. A massless gauge field theory may pose additional problems even if it taken as the limit of a massive gauge field theory.  The exact features of the unitary gauge in all orders of perturbation theory is an open issue that deserves more exploration.  However in principle one may assume at any time some particular gauge and the results that may be obtained in it. Thus the arguments of this work that lead to the quantization of charge remain pertinent.

In conclusion we proved that the quantization of charge in electrodynamics is an intrinsic property of the theory if a justified quantum consistency condition is required. Applied to the standard model hypercharge gauge boson our criterion together with the cancellation of the anomalies lead to the correct hypercharge and consequently charge assignment for all the fermions. As opposed to this, note that by itself  the standard model does not require the known hypercharge assignment and this is based on experimental observation.

Our method can be used for any gauge abelian sector of a more comprehensive theory  as a check of consistency.

\end{document}